\documentclass[aps,prb,twocolumn,superscriptaddress,showpacs]{revtex4}
\usepackage{amsmath}
\usepackage{bm}
\usepackage{graphicx}

\begin{document}

\title{Kondo effect in a double quantum-dot molecule under the effect of an
electric and magnetic field}
\author{Gustavo. A. Lara}
\affiliation{Departamento de F\'{\i }sica, Universidad de
Antofagasta,Casilla 170, Antofagasta, Chile.}
\author{Pedro A. Orellana}
\affiliation{Departamento de F\'{\i }sica, Universidad Cat\'{o}lica del
Norte,Casilla 1280,Antofagasta, Chile.}
\author{Enrique V. Anda}
\affiliation{Departamento de F\'{\i }sica, P. U. Cat\'{o}lica do Rio de
Janeiro,C.P. 38071-970,Rio de Janeiro, RJ, Brazil.}

\begin{abstract}
Electron tunneling through a double quantum dot molecule, in the Kondo
regime, under the effect of a magnetic field and an applied voltage, is
studied. This system possesses a complex response to the applied fields
characterized by a tristable solution for the conductance. The different
nature of the solutions are studied in and out thermodynamical equilibrium.
It is shown that the interdot coupling and the fields can be used to control
the region of multistability. The mean-field slave-boson formalism is used
to obtain the solution of the problem.
\end{abstract}


\maketitle

The electron tunneling through a double-quantum-dot (DQD) molecule in the
Kondo regime has received much attention in the last years.\cite
{jeong,aguado,georges,busser,aono,eto,izumida,ivanov,orellana} Recently
Jeong et al.\cite{jeong} reported the observation of a coherent Kondo effect
in a double quantum dot, which is a direct experimental realization of a
two-impurity Anderson model.\cite
{aguado,georges,busser,aono,eto,izumida,ivanov} The strength of the coupling
between the QD's determines the character of the molecular electronic states
and the transport properties of the artificial molecule. In the tunneling
regime, the states are extended across the entire system and can be
constructed from a coherent state based on the bonding or anti-bonding
levels of the quantum dots. These many-body molecular states are observed in
a transport experiment as a differential conductance of two peaks, a
consequence of the splitting of the Kondo resonance.

Recently Aono and Eto studied theoretically the problem of a DQD in
thermodynamical equilibrium, under the effect of a magnetic field oriented parallel to the current direction. \cite{eto}%
They calculated the magnetoconductance and found a suddenly drop of the
conductance as a function of the field, when $t_{c}>\Gamma,$ where $t_{c}$
is the interdot interaction and $\Gamma $ the resonant width of the dot
levels due to coupling among them and the contacts. Some aspects of the
non-equilibrium transport properties of a DQD in the Kondo regime, under the
effect of an applied voltage, has recently been studied \cite{orellana}.
When the inter-dot coupling is greater than the level broadening and there
is an external voltage applied to the system, the conductance shows a
tristable behavior.

In this work we analyze the physics derived from the complex behavior that
characterizes the electron tunneling through a DQD molecule in the Kondo
regime under the simultaneous effect of a magnetic and an electric field. We
show that the discontinuity of the conductance found by Aono and Eto,\cite
{eto} increasing the magnetic field when the system is in equilibrium, is a
result of a tristable behavior of the magnetoconductance. This results is
similar to the behavior the system has when it is out of equilibrium, due to
the application of an external voltage.\cite{orellana}

We consider a DQD connected in series attached to two leads. The physics
associated to this artificial molecule electrically connected can be readily
understood in terms of two impurities Anderson Hamiltonian where the
impurities located in the site $0$ and $1$ are the quantum dots. The leads
are represented by a one dimensional tight-binding Hamiltonian connecting
the dots to two particle reservoirs characterized by Fermi levels $\mu _{L}$
and $\mu _{R}$ respectively. The left (right) dot is connected to the left
(right) lead by the hopping $V_{L(R)}$ and the dots are connected among
themselves by the interdot tunneling coupling $t_{c}.$ Each dot $\alpha $
has a single energy $\varepsilon _{\alpha }(\alpha =0,1)$. Under an applied
magnetic field the energy levels of the dots are split by the Zeeman effect
into $\varepsilon _{\alpha }\pm \sigma g\mu _{B}B$. We adopt the two-fold
degenerate Anderson Hamiltonian in the limit $U\rightarrow \infty ,$ that is
diagonalized using the mean-field slave-boson formalism. \cite
{barnes,coleman,read} The double occupancy in each dot is forbidden and the
inter-dot Coulomb interaction neglected. We introduce the slave-boson
operator $b_{\alpha }^{\dagger }$ that creates an empty state and a fermion
operator $f_{\alpha \sigma }$ that annihilates a single occupied state with
spin $\sigma $. To eliminate the possibility of double occupancy we impose
the constraint $Q_{\alpha }\equiv \sum\nolimits_{\sigma }f_{\alpha \sigma
}^{\dagger }f_{\alpha \sigma }+b_{\alpha }^{\dagger }b_{\alpha }=1$. The
annihilation operator of an electron in the dot $\alpha $ is $c_{\alpha
\sigma }=b_{\alpha }^{\dagger }f_{\alpha \sigma }$. In the mean field
approximation the bosonic operators $b_{\alpha }^{\dagger }$ and $b_{\alpha
} $ are replaced by their expectation values, $\left\langle b_{\alpha
}\right\rangle =\widetilde{b}_{\alpha }\sqrt{2}=\left\langle b_{\alpha
}^{\dagger }\right\rangle =\widetilde{b}_{\alpha }^{\dagger }\sqrt{2}.$
Hence the Hamiltonian of the DQD molecule connected to leads plus
constraints is written as,

\begin{widetext}
\begin{eqnarray}
H &=&H_{lead}+\sum\limits_{\alpha =0,1,\sigma }(\widetilde{\varepsilon }%
_{\alpha }+\sigma g\mu _{B}B)f_{\alpha \sigma }^{\dagger }f_{\alpha \sigma }
+\widetilde{V}_{L}\sum_{\sigma }(c_{-1\sigma }^{\dagger }f_{0\sigma
}+f_{0\sigma }^{\dagger }c_{-1\sigma })+\widetilde{V}_{R}\sum_{\sigma
}(f_{1\sigma }^{\dagger }c_{2\sigma }+c_{2\sigma }^{\dagger }f_{1\sigma }) \nonumber \\
&+&\widetilde{t}_{c}\sum_{\sigma }(f_{0\sigma }^{\dagger }f_{1\sigma
}+f_{1\sigma }^{\dagger }f_{0\sigma })+\sum\limits_{\alpha =0,1}\lambda
_{\alpha }(\widetilde{b}_{\alpha }^{\dagger }\widetilde{b}_{\alpha }-1).
\end{eqnarray}
\end{widetext}

\noindent where $\widetilde{\varepsilon }_{\alpha }=\varepsilon _{0}+\lambda
_{\alpha }$, $\widetilde{V}_{L}=V_{L}\widetilde{b}_{0}$, $\widetilde{V}%
_{R}=V_{R}\widetilde{b}_{1}$, $\widetilde{t}_{c}=t_{c}\widetilde{b}_{0}%
\widetilde{b}_{1},$ the $\lambda _{\alpha }$ are Lagrangian multipliers,
which guarantee the constraint conditions on the $Q_{\alpha }$ .

\bigskip \noindent The first term on the right-hand side of Eq.1 represents
the electrons in the left and right leads;

\begin{equation}
H_{lead}=\sum_{\sigma ,i\neq 0,1}\varepsilon _{i}n_{i\sigma }+t\sum_{<ij\neq
0,1>\sigma }c_{i\sigma }^{\dagger }c_{j\sigma }.
\end{equation}

\noindent where the operator $c_{i\sigma }^{\dagger }$ creates an electron
in the site $i$ with spin $\sigma $, $\varepsilon _{i}$ is the site energy
and $t$ is first-neighbor hopping in the leads. Here $H_{lead}$ is a
free-particle Hamiltonian with eigenfunctions of the Bloch type,

\begin{equation}
\left| k,\sigma \right\rangle =\sum_{j}e^{ikja}\left| j,\sigma \right\rangle
,
\end{equation}

\noindent where $\left| k,\sigma \right\rangle $ is the momentum eigenstate
with spin $\sigma $ and $\left| j\right\rangle $ is a Wannier state
localized at site $j $ of spin $\sigma $. The dispersion relation associated
with these Bloch states reads

\begin{equation}
\varepsilon =2t\cos (kd).
\end{equation}
The model has an energy band, extending from $-2t$ to $+2t$, with the first
Brillouin zone expanding the $k$ interval $[-\pi /d,\pi /d]$.

The stationary state of the complete Hamiltonian $H$ with energy $%
\varepsilon _{k}$ can be written as 
\begin{equation}
\left| \psi _{k\sigma }\right\rangle =\sum_{j}a_{j\sigma }^{k}\left|
j,\sigma \right\rangle ,
\end{equation}
\noindent where the coefficients $a_{i\sigma }^{k}$ are given by,

\begin{equation}
a_{j\sigma }^{k}=<j,\sigma |\psi _{k\sigma }>.
\end{equation}

We obtain the following eigenvalue equations for the Wannier amplitudes $%
a_{j\sigma }^{k}$

\begin{eqnarray}
\varepsilon a_{j,\sigma }^{k} &=&\left\langle j,\sigma \left| H\right| \psi
_{k\sigma }\right\rangle  \nonumber \\
\varepsilon a_{j,\sigma }^{k} &=&\varepsilon _{j}a_{j,\sigma
}^{k}+t(a_{j-1,\sigma }^{k}+a_{j+1,\sigma }^{k})\;\;\;\;(j\neq -1,0,1,2), 
\nonumber \\
\varepsilon a_{-1(2),\sigma }^{k} &=&\varepsilon _{-1(2)}a_{-2(2),\sigma
}^{k}+\widetilde{V}_{L(R)}a_{0(1),\sigma }^{k}+ta_{-2(3),\sigma }^{k}, 
\nonumber \\
\varepsilon a_{0(1),\sigma }^{k} &=&(\widetilde{\varepsilon }_{0(1)}+\sigma
g\mu _{B}B)a_{0(1),\sigma }^{k}+\widetilde{V}_{L(R)}a_{-1(2),\sigma }^{k} 
\nonumber \\
&+&\widetilde{t}_{c}a_{1(0),\sigma }^{k},
\end{eqnarray}

\noindent where $a_{j,\sigma }^{k}$ is the probability amplitude to find the
electron in site $j$ and state $k$ with spin $\sigma $.

The four parameters ($\widetilde{b}_{0}$, $\widetilde{b}_{1}$, $\lambda _{0}$%
, $\lambda _{1}$) are determined minimizing the expectation value of the
Hamiltonian (1). They satisfy the set of equations: 
\begin{widetext}
\begin{eqnarray}
\widetilde{b}_{0(1)}^{2}+\frac{1}{2}\sum\limits_{k,\sigma }\left|
a_{0(1),\sigma }^{k}\right| ^{2} =
\frac{1}{2},  \nonumber \\
\frac{\widetilde{V}_{L(R)}}{2}\sum\limits_{k,\sigma }{%
\mathop{\rm Re}%
}(a_{-1(2),\sigma }^{k*}a_{0(1),\sigma }^{k}) +
\frac{\widetilde{t}_{c}}{2}\sum\limits_{k,\sigma }{%
\mathop{\rm Re}%
}(a_{1(0),\sigma }^{k*}a_{0(1),\sigma }^{k})+\widetilde{\lambda }_{0(1)}%
\widetilde{b}_{L(R)}^{2} =0.
\end{eqnarray}
\end{widetext}

\noindent The sum over the wave vector $k$ cover all the occupied states.
The resulting equations are nonlinear due to the renormalization of the
localized levels in the dots, the interdot coupling tunneling and the
coupling tunneling between the quantum dots and the leads.

In order to study the solutions of equations (7) and (8) we assume a plane
wave incident from one lead with an amplitude $I$ and with a partial
reflection amplitude $R$. The waveform at the other lead is a simple plane
wave with intensity given by the transmission amplitude $T$. Taking this to
be the solution of the system at infinite, for positive $k$, we can write 
\begin{eqnarray}
a_{j\sigma }^{k} &=&I_{\sigma }^{+}e^{ikdj}+R_{\sigma }^{+}e^{-ikdj}\text{ }%
,\quad j<0  \nonumber \\
a_{j\sigma }^{k} &=&T_{\sigma }^{+}e^{ik_{R}dj},\quad \qquad \qquad \text{ }%
j>1
\end{eqnarray}

\noindent and those with negative $k$

\begin{eqnarray}
a_{j\sigma }^{-k} &=&I_{\sigma }^{-}e^{-ikdj}+R_{\sigma }^{-}e^{ikdj},\qquad
\quad j>1,  \nonumber \\
a_{j\sigma }^{-k} &=&T_{\sigma }^{\_}e^{-ik_{L}dj},\qquad \qquad \qquad
\;j<0.
\end{eqnarray}
These functions describe free electrons approaching the scattering potential
from $j=-\infty $ and $j=+\infty ,$ respectively, with transmission and
reflections amplitudes $T_{\sigma }^{+}$ and $R_{\sigma }^{+}$ for $k>0$ and 
$T_{\sigma }^{\_}$ and $R_{\sigma }^{-}$ for $k<0$.

The solution of equations (7) can be obtained through an adequate iteration
from right to left for $k>0$, and from left to right for $k<0$. For a given
transmitted amplitude, the associated reflected and incident amplitudes may
be determined by matching the iterated function to the proper plane wave at
the left lead for $k>0$ and at the right lead for $k<0.$ The transmission
amplitude $t_{\sigma }=T_{\sigma }^{+}/I_{\sigma }^{+}$ and $\widetilde{t}%
_{\sigma }=T_{\sigma }^{\_}/I_{\sigma }^{-}$ are obtained using an iterative
procedure. The total transmission probability with a spin $\sigma $ and
energy $\varepsilon =$ $2t\cos (kd)$ given by $T(\varepsilon
)=\sum\limits_{\sigma =\uparrow ,\downarrow }\left| t_{\sigma }(\varepsilon
)\right| ^{2}.$ Then the linear conductance at zero temperature is written
as,

\begin{equation}
G=\frac{2e^{2}}{h}T.
\end{equation}

The equations (7) and (8) are nonlinear and require a self-consistent
solution that is obtained using a conjugate gradient algorithm. In this
work, we are interested in analyzing all the stationary solutions for the
conductance. As $G(B)$and $B(G)$ are multivalued functions, the multiple
solutions for the quantities $\widetilde{b}_{0}$, $\widetilde{b}_{1}$, $%
\lambda _{0}$, $\lambda _{1}$ are obtained by fixing the magnetic field $B$
and obtaining the value of the conductance $G$ that corresponds to the
solution of the equations of motion of the system. This method permits us to
obtain the complete $G-B$ curve.

We study first a model which consists of two leads equally connected $%
(V_{L}=V_{R}=V_{0})$ to both sides of a DQD dots in equilibrium with
chemical potential $\mu _{L}=\mu _{R}=0$, $t_{\text{ }}=30\;\Gamma _{0}$, $%
V_{0}=5.48\;\Gamma _{0}$, $\varepsilon _{0}=\varepsilon _{1}=-3.5\;\Gamma
_{0}$ (Kondo regime with $T_{K}\approx 10^{-3}\;\Gamma _{0}$ with $\Gamma
_{0}=2\pi V_{0}^{2}\rho (0)$).

\begin{figure}[h]
\centerline{\includegraphics[scale=0.28,angle=270]{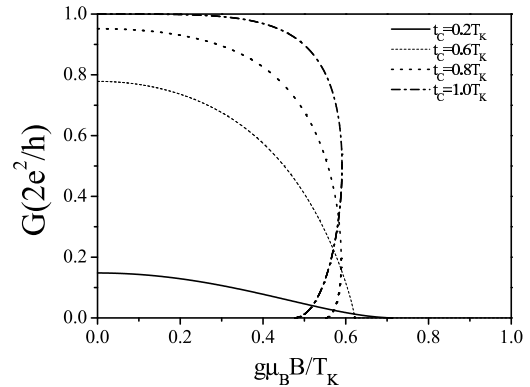}}
\caption{$G$-$B$ curve for a)$t_{c}=0.2\Gamma_{0}$, b)$t_{c}=0.6\Gamma_{0}$,
c)$t_{c}=0.8\Gamma_{0}$ and d) $tc=\Gamma_{0}$}
\end{figure}

Fig 1 shows the linear conductance versus the applied magnetic field for
different values of $t_{c}$ with $t_{c}\leq \Gamma _{0}$. For small values
of $t_{c}$, $t_{c}\ll \Gamma _{0}$ our results agrees with those obtained by
Aono and Eto. The magnetoconductance is negative in this region. However for
greater values of $t_{c}$, $t_{c}\sim \Gamma _{0}$ another behavior of the
magnetoconductance starts. Within a range of values of the magnetic field
there is a tristability behavior with three stationary solutions of the
conductance for each value of the magnetic field. Fig 2 shows the results
for $t_{c}>\Gamma _{0}.$ As in the results found by Aono and Eto, at a
critical value $B_{c\uparrow ,}$ there would be a discontinuous reduction of
the conductance when $B$ is increased and other $B_{c\downarrow
}(B_{c\uparrow }>B_{c\downarrow })$ where the conductance would rise
discontinuously, when $B$ is reduced. Using this procedure, the interior of
the tristability remains inaccessible to the experiment. However, the
internal shape could eventually be reached, as it is for the case of double
barrier heterostructrures, if it were possible to employ a load line in the
experimental measurement to go along all the points of the $G-B$ curve. From
this point of view the electrons goes along a nonlinear potential, that is
magnetic field dependent. In this case the magnetic field plays the role
that the external electric potential plays in the case of a standard double
barrier heterostructure, where the current has as well a multistable
behavior.\cite{eaves}

\begin{figure}[h]

\centerline{\includegraphics[scale=0.28,angle=270]{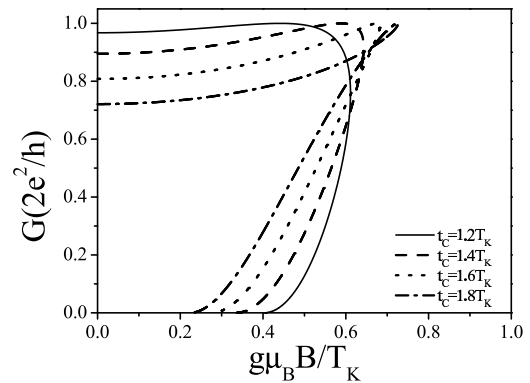}}
\caption{Linear conductance $G$ versus the Zeeman splitting $g\mu_B B$ for a)%
$t_{c}=1.0\Gamma_{0}$, b)$t_{c}=1.2\Gamma_{0}$, c)$t_{c}=1.6\Gamma_{0}$ and
d) $tc=1.8\Gamma_{0}$}
\end{figure}

Increasing the magnetic field the tristability probably arises from a
many-body ground state that changes its character due to a energy crossing.
It is a singlet of the Kondo type when there is no applied magnetic field,
while for a critical field onwards the ground state results to be the
component of the triplet wave function, (two parallel spins one at each dot)
with spin projection parallel to the magnetic field. In the parameter space
this two ground states have their own basin of attraction constituting
metastable states when they loose the condition of been the ground state of
the system. As soon as the metastability is lost increasing or decreasing
the field, the system abruptly relaxes to the ground state producing a
discontinuous behavior of the current.\cite{anda1}

It is very interesting to study these non-linear effect when the system is
under the simultaneous action of an external magnetic and electric field. In
this case the DQD is out of equilibrium as it operates within two chemical
potentials $\mu _{L}=V/2$ and $\mu _{R}=-V/2$ defined by the applied dc bias
voltage $V$. Imposing a fixed value of the magnetic field, once the
amplitudes $a_{j,\sigma }^{k}$ are known, the current can be numerically
obtained from,

\begin{equation}
J=\frac{2e}{\hbar }\tilde{V}_{L}\sum\limits_{k,\sigma }{%
\mathop{\rm Im}%
}\{a_{-1,\sigma }^{k*}a_{0,\sigma }^{k}\}.
\end{equation}

\begin{figure}[h]
\centerline{\includegraphics[scale=0.28,angle=270]{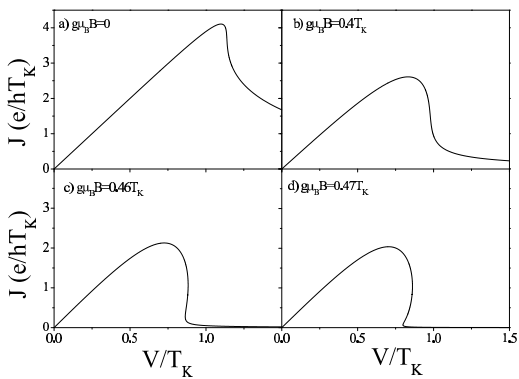}}
\caption{$J$-$V$ characterisitic curve for $t_{c}=\Gamma_{0}$ for various
values of the Zeeman splitting $g\mu_B B$. a)$g\mu_B B=0.0$ , b)$g\mu_B
B=0.40T_{K}$, c)$g\mu_B B=0.46T_{K}$ and d) $g\mu_B B=0.47T_{K}$}
\end{figure}

We study first the problem when $t_{c}=\Gamma $. For this value of $t_{c}$,
in the absence of an external magnetic field, the system is at the threshold
of having a manifestation of the non-linearity produced by correlation. For
a regime in which $g\mu _{B}B\ll T_{K}$, figure 3 shows that the magnetic
field reduces the height of the peak and valley of the $J-V$ characteristic
curve. However, for larger magnetic fields, $g\mu _{B}B<T_{K}$, a
tristability behavior of the current is promoted. For greater values of $%
t_{c}$ ($t_{c}=1.4\Gamma $), where the system has already a highly non-linear
behavior in the absence of magnetic field, the shape of the tristability is
highly dependent on $B$. As shown in figure 4, here as well, a remarkable
increase of the peak to valley ratio is obtained in the regime $g\mu
_{B}B<T_{K}$. This ratio goes from about $20$ at zero magnetic field to
about $2000$ at $g\mu _{B}B=0.333T_{K}$. These phenomena are a result of
the reduction and eventual elimination of the Kondo correlation when the
magnetic field is of the order of the Kondo temperature. The Zeeman effect
destroys the coherent behavior of the two QD's produced by the molecular
Kondo coupling, the electrons localize at each dot occupying levels that are
mismatched by the applied potential and consequently the current goes to
zero.

\begin{figure}[h]
\centerline{\includegraphics[scale=0.28,angle=270]{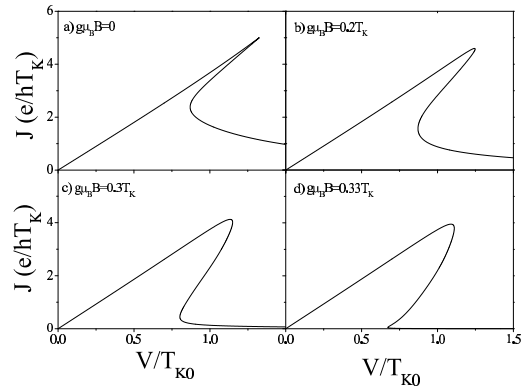}}
\caption{$J$-$V$ characteristic curve for $t_{c}=1.4\Gamma_{0}$ for various
values of the magnetic field. a)$g\mu_B B=0$ , b)$g\mu_B B=0.2T_{K}$, c)$%
g\mu_B B=0.3T_{K}$ and d) $g\mu_B B= 0.33T_{K}$}
\end{figure}

In non-equilibrium the nonlinear behavior described above has similarities
with the well-known bistability present in three dimensional resonant
tunneling double barrier systems and multistabilities in doped
superlattices. \cite{goldman,anda,orellana,prengel,kastrup,aguado2} In
three dimensional tunneling double barrier systems, the Coulomb interaction
pins the renormalized energy level at the well when the applied voltage $V$
is augmented maintaining the system at resonance. However, when $V$ goes
above a critical value, the abrupt leakage of the charge accumulated in the
well takes the system out of resonance. For doped superlattices, at large
carrier densities, the internal potential profile produced by the applied
voltage breaks up into domains giving rise to a multistable behavior in the
current.\cite{prengel,kastrup,aguado2} In both systems, the non-linear
deformation of the potential profile due to the electron-electron
interaction is responsible for the phenomenon. Although in our case the
Kondo effect is a strong many body effect of a different nature in
comparison with the previous systems, the multistability is as well a
consequence of the non-linear potential profile deformations that the
electrons suffer when the external potential is modified. 

In conclusion, we have investigated the magnetoconductance in a DQD molecule
in the Kondo regime. We have shown that the magnetoconductance has a
tristable behavior depending of the ratio of $t_{c}/\Gamma $ and the value
of the magnetic field that controls the non-linearity of the system. This evidence was
obtained using a mean-field approach that neglects fluctuations, appropriate
for large spins as it is the first term of an $1/N$ expansion, where $N$ is
the spin dimension. Some experiments have shown\cite{sasaki} that it is
possible to design dots with a configuration that, due to the exchange
interaction, the parallel spin coupling following Hund's rule gives rise to
a total dot spin $S_{t}>1/2$. For this cases our results are more reliable
as fluctuations are less important. Although we believe that our results are
robust against them, a study that goes beyond mean field is necessary to
confirm our conclusions. 

Work supported by grants Milenio ICM P99-135-F, FONDECYT Grants 1020269 and 7020269 and Red IX.E 
"Nanoestructuras para la Micro y Optoelectronica Sub-Programa IX "Microelectronica" Programa CYTED.


\begin{thebibliography}{99}
\bibitem{jeong}  H. Jeong, A.M. Chang and M.R. Melloch, Science \textbf{293}
2221 (2001).

\bibitem{aguado}  R.Aguado and D.C. Langreth Phys.Rev.Lett. \textbf{85},1946
(2000).

\bibitem{georges}  Antoine Georges and Yigal Meir, Phys. Rev. Lett. \textbf{%
82}, 3508 (1999).

\bibitem{busser}  C.A. Busser, E.V. Anda, A.L. Lima, M. A.Davidovich and G.
Chiappe, Phys. Rev. B \textbf{62}, 9907 (2000).

\bibitem{aono}  Tomosuke Aono and Mikio Eto, Phys. Rev. B \textbf{63},125327
(2001).

\bibitem{eto}  Tomosuke Aono and Mikio Eto, Phys. Rev. B \textbf{64},073307
(2001).

\bibitem{izumida}  Wataru Izumida and Osamu Sakai, Phys. Rev. B \textbf{62}
10260 (2000).

\bibitem{ivanov}  T. Ivanov Europhys. Lett. \textbf{40}, 183 (1997).

\bibitem{orellana}  Pedro A. Orellana, G.A. Lara, Enrique V. Anda, Phys.
Rev.B \textbf{65} 155317 (2002),Gustavo A. Lara, Pedro A. Orellana, Enrique
V. Anda, Sol. State Comm. \textbf{125} 165 (2003).

\bibitem{barnes}  S.E. Barnes, J. Phys. F \textbf{6}, 1375 (1976); \textbf{7}%
, 2637 (1977).

\bibitem{coleman}  P. Coleman, Phys. Rev. B \textbf{29}, 3035 (1984); 
\textbf{35}, 5072 (1987).

\bibitem{read}  N. Read and D.M. Newns, J. Phys. C, 3273 (1983); Adv. Phys. 
\textbf{36}, 799 (1988).

\bibitem{eaves}  A.D Martin, M.K.F. Lerch, P.E. Simmonds and E. Eaves, Appl.
Phys. Lett. 64, 1248 (1994).

\bibitem{goldman}  V.J. Goldman, D.C.Tsui and J.E. Cunningham, Phys. Rev.
Lett. \textbf{58}, 1256 (1987).

\bibitem{prengel}  F. Prengel, A. Wacker, and E. Scholl, Phys. Rev. B 
\textbf{50}, 1705 (1994).

\bibitem{kastrup}  J. Kastrup,R. Hey, K. H. Ploog, H. T. Grahn, L. L.
Bonilla, M. Kindelan, M. Moscoso, A. Wacker, and J. Gal\'{a}n, Phys. Rev. B 
\textbf{55}, 2476 (1997).

\bibitem{aguado2}  R. Aguado,G. Platero, M. Moscoso, and L. L. Bonilla,
Phys. Rev. B \textbf{55}, R16053 (1997).

\bibitem{anda}  P.L. Pernas, F. Flores and E.V. Anda, Phys. Rev. B \textbf{47%
}, 4779 (1993).

\bibitem{anda1}  E. V. Anda ,V. Ferrari and G. Chiappe,
J.Phys.:Condens.Matter \textbf{9}, 1095 (1997).

\bibitem{jones}  B.A. Jones, B.G. Kotliar, A. J. Millis, Phys. Rev. B 
\textbf{39}, R3415 (1989).

\bibitem{sasaki}  S.Sasaki, S. De Franceschi, J.M. Elzerman, W.G. van de Wiel,
M. Eto, S. Tarucha and L. P. Kowenhoven, Nature \textbf{405, }764 (2000).
\end{thebibliography}
\end{document}